# Vibrational Modes and Instabilities of a Dust Particle Pair in a Complex Plasma

K. Qiao, L. S. Matthews, and T. W. Hyde, *Member, IEEE*

*Abstract*— Vibrational modes and instabilities of a dust particle pair in a terrestrial laboratory complex plasma are investigated employing an analytical method whereby the plasma wakefield induced by an external electric field is modeled using an image charge method. It is found that for both horizontally and vertically aligned dust particle pairs in equilibrium, four normal modes exist. Variations of the confinement parameters cause a single type of instability in the horizontal pair and two types of instabilities in the vertical pair.

## I. Introduction

PAIR interactions are of fundamental interest in research on many-body systems. One such system, which is relatively new and exhibits many unique properties, is complex plasma, where micron-sized dust particles are immersed in an electron/ion low temperature plasma, charging negatively to $10^3$-$10^4$ elementary charges. The pair interaction potential between these dust particles is primarily determined by the local plasma environment surrounding them. For cases without an external electric field, the local plasma is symmetric around the dust particles producing a screened Coulomb, or Debye-Hückel, potential [1], In the quasi-neutral bulk plasma, where ions are in rough equilibrium with the neutrals and the electron temperature is usually a few eV, the screening length is approximately equal to the ion Debye length. However, in the sheath region where dust particles levitate in most experimental complex plasmas on Earth, the ion flow velocity is such that the mean kinetic energy of the ions can equal or exceed that of the bulk electrons [2]. In this case, experimental measurements have shown the screening length to be roughly equal to the bulk electron Debye length [1]; thus, screening is often attributed to the electrons rather than the ions [3, 4]. Due to the complexity of the plasma sheath, the origin of this screening remains an open question. Many of the properties exhibited by the two-dimensional (2D) plasma crystal which can form under these circumstances, such as its hexagonal structure [5]-[7], ability to support longitudinal [8] transverse [9] and out-of-plane dust lattice waves [10,11], dust acoustic waves [12], and so on, can be attributed to the characteristics provided by the screened Coulomb potential.

Any external electric field, such as that found in experimental complex plasmas on Earth [5]-[7] and some electrically controlled micro-gravity systems [13], will affect the plasma reshaping the interaction potential around the dust particles. As mentioned, dust particles in laboratory complex plasmas typically levitate in the sheath region above a powered lower electrode within an rf discharge plasma. In this region, the electric field points toward the lower electrode creating an ion flow in this direction. The corresponding potential field around the dust is commonly referred to as the wakefield. The wakefield has been studied extensively both theoretically [2-4, 14-18] and experimentally [19]-[23] due to its importance to the overall structural development and dynamics of complex plasmas. Experiments have found that the wakefield is attractive and nonreciprocal [19, 20], which may be the cause for the vertically aligned structure of the resulting plasma crystals [2-4, 14-23].

Theoretically the wakefield has been attributed to either a space-charge effect, due to an increased density of ions below the dust particle [2-4, 14-18], or the ion drag force from the horizontal velocity component of the ion flow [24]. These two effects have the same order of magnitude [24] leaving the mechanism causing the wakefield an open question. For this paper, the wakefield will be assumed to be a space-charge effect, due to an increased density of ions below the dust particle.

Manuscript received June 30, 2009. This work was supported in part by the National Science Foundation under Grant BS123456.

Ke Qiao, Lorin Matthews and Truell Hyde are with the Center for Astrophysics, Space Physics & Engineering Research, Baylor University, Waco, TX USA (Truell Hyde: 254-710-3763; fax: 254-710-7309; e-mail: Truell_Hyde@Baylor.edu).

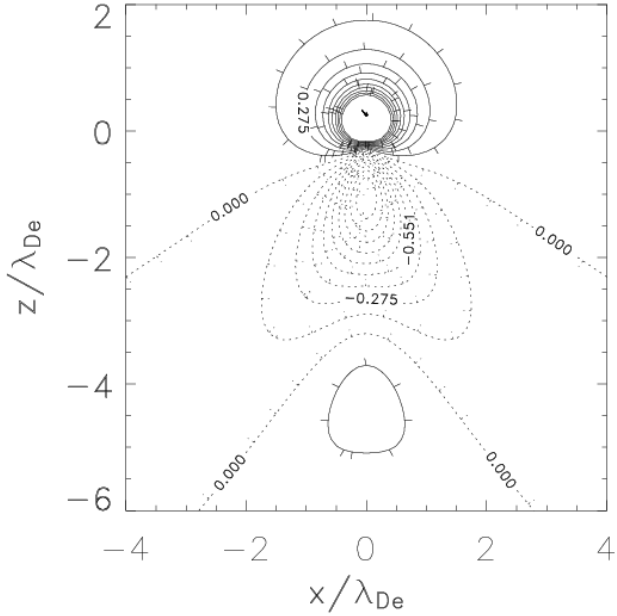

Fig. 1. Contour plot of the potential around a dust particle (centered at the origin) obtained using linear response theory. Dashed contours represent negative potential surfaces with tick marks indicating decreasing direction. Taken from [14].

The most accurate theoretical model of the wakefield obtains the potential using linear response theory (Fig. 1) [2, 14, 15]. Unfortunately, the mathematical representation of this potential is very complex. A much simpler model is the image charge model, where one assumes the wakefield around a dust particle with negative charge $Q$, takes the form of the potential created by a positive point charge $q$ located at a distance $l$ beneath the dust grain (Fig. 2) [16]-[18], [21]. By choosing appropriate values for $q$ and $l$, linear response results can be well-fit by employing a dipole-monopole potential in the region of space near the first minimum in the wake potential. (Fig. 1) (This is the region where dust particles are typically found in experiment.)

The simplest system that can be used to investigate the particle interaction potential including the effect of the wakefield is the dust particle pair. The non-reciprocity of the wakefield effect on this interaction provides a host of rich physics phenomena for investigation. One example is the particle transition from horizontal to vertical alignment and back; this effect has been studied experimentally [21-23] and shown to display a hysteresis effect. Theoretically this system can be analyzed using a Hamiltonian or perturbation theory [14-18].

Following a Hamiltonian approach typically involves introducing an effective Hamiltonian even though this system is inherently non-Hamiltonian [14, 16, 17]. As such, the stability of the system is best analyzed by mapping the effective potential using the linear response potential model [14] or the image charge model [16, 17]. (Hamiltonian dynamics may be applicable when the system is in static equilibrium where the energy dissipation is equal to the energy gain by the system [17]. However, this is still under debate.)

Employing perturbation theory involves solving Newton's equations for the oscillation frequencies employing a linear perturbation method. This technique has been used by Vladimirov and Samarian [15, 17] adopting the linear response results (on the z-axis connecting a vertically aligned pair). However, the horizontal and vertical motions were investigated separately assuming decoupling between them, and for horizontally aligned pairs only the Yukawa potential interaction was considered.

In each of the cases mentioned above, the research focused on the transition between horizontal and vertical alignment and the instabilities which can create them. The wake dependence of the oscillation spectrum of particle pairs in equilibrium has not received nearly as much attention. In this research, the oscillation mode spectra for both horizontally and vertically aligned dust particle pairs under typical laboratory plasma conditions are investigated using an analytical method based on the image charge model and including the effects of the ion flow. This method starts with Newton's equations, thus treating the system as non-Hamiltonian from the outset. Oscillation modes are obtained by solving the resulting set of equations, thus avoiding the assumption of mode decoupling and at the same time naturally including the motion of the center of mass (COM). Instabilities are identified as natural results for oscillation modes producing negative frequencies. The dependence of these on the wakefield is obtained, where the wakefield is represented by $q$ and $l$, the image charge magnitude and distance beneath the dust particle as described above.

The analytical method is described in Section II. The mode spectra, instabilities and the relationship between them and the image charge is discussed for horizontally aligned dust particle pairs in Section III and for vertically aligned dust particle pairs in Section IV. A summary of results is given in Section V.

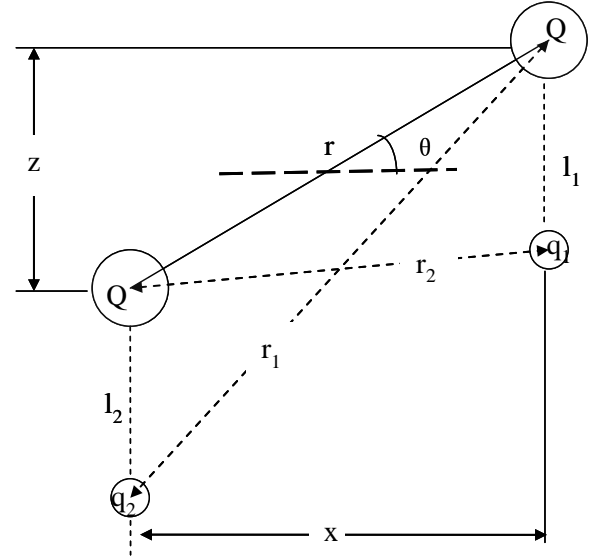

Fig. 2. A dust pair system consisting of particles having equal charge, $Q$, and inclined at an angle $\theta$ to the horizontal. Image charges, $q_1$ and $q_2$, are also shown.

## II. ANALYTICAL METHOD

Two dust particles comprising a pair within a laboratory complex plasma are confined by a parabolic potential well in both the horizontal [25] and vertical directions [26],

$$U_{ext}(x,y,z) = \frac{1}{2}\left[c_x(x^2+y^2)+c_z z^2\right], \quad (1)$$

where $x$, $y$, $z$ are representative particle coordinates, and $c_x$ and $c_z$ are the horizontal and vertical confinement parameters defined as the spatial derivative of the external electric field, $dE/dx$ and $dE/dz$.

The experimental environment considered in this research is similar to that discussed in [21]: a low-density plasma with a screening length (experimentally measured to be on the same order as the bulk electron Debye length) much larger than the particle separation distance. Thus, the interaction between the dust particle and any image charge can be treated using an unscreened Coulomb potential.

The pair interaction between any two dust particles having equal charges $Q$ and masses $m$ where the surrounding plasma distribution is modified due to a DC electric field can be modeled by assuming there is an image charge $q_i$ (where $i = 1$ or 2 represents the dust particle index), located a distance $l_i$ below the $i$th dust particle (Fig. 2). The resulting pair interaction now includes the dust-dust interaction, the interaction between dust particle 1 and image particle 2, and the interaction between dust particle 2 and image particle 1. This interaction is non-Hamiltonian [15], thus the net force on the pair due to this interparticle interaction is not zero. Therefore the center of mass (COM) of the dust particle pair at equilibrium will no longer coincide with the minimum of the potential well, in agreement with the results shown in [14]. As such, any reliable theoretical investigation must start from Newton's equations taking into account COM motion, as shown below.

$$x\left(\frac{1}{r^3} - \frac{q_2}{r_1^3}\right) - \frac{1}{2}c_x(x+2x_c) = m\frac{d^2x_1}{dt^2} \quad (2)$$

$$x\left(\frac{1}{r^3} - \frac{q_1}{r_2^3}\right) - \frac{1}{2}c_x(x-2x_c) = -m\frac{d^2x_2}{dt^2} \quad (3)$$

$$\left(\frac{z}{r^3} - \frac{q_2 z_1}{r_1^3}\right) - \frac{1}{2}c_z(z+2z_c) = m\frac{d^2z_1}{dt^2} \quad (4)$$

$$\left(\frac{z}{r^3} - \frac{q_1 z_2}{r_2^3}\right) - \frac{1}{2}c_z(z-2z_c) = -m\frac{d^2z_2}{dt^2} \quad (5)$$

The horizontal and vertical positions of the $i$th particle are denoted as $x_i$ and $z_i$, with $x = x_1 - x_2$, $z = z_1 - z_2$, $x_c = (x_1 + x_2)/2$, and $z_c = (z_1 + z_2)/2$ being the relative and COM positions in the horizontal and vertical directions, respectively. The interparticle distance is given by $r$, with the distance between the $i$th particle and image charge $q_j$ given by $r_i$. The projections of $r_1$ and $r_2$ on the vertical axis are given by $z_1 = z + l_2$ and $z_2 = z - l_1$. (All variables in the above are dimensionless through normalization by $Q$, the unit charge, and $l_1$, the unit length. This normalization will apply throughout the paper unless specifically stated otherwise.) The confinement parameters, $c_x$ and $c_z$ are normalized by $c_0 \equiv Q/l_1^3$. A representative dust pair with each of these parameters indicated is shown in Fig. 2.

Combining Equations 2, 3, 4 and 5 yields four equations with $x$, $z$, $x_c$ and $z_c$ as general coordinates:

$$x\left[\frac{2}{r^3} - \left(\frac{q_2}{r_1^3} + \frac{q_1}{r_2^3}\right)\right] - c_x x = m\frac{d^2x}{dt^2} \quad (6)$$

$$\left[\frac{2z}{r^3} - \left(\frac{q_2 z_1}{r_1^3} + \frac{q_1 z_2}{r_2^3}\right)\right] - c_z z = m\frac{d^2z}{dt^2} \quad (7)$$

$$-c_x x_c - \frac{1}{2}x\left(\frac{q_2}{r_1^3} - \frac{q_1}{r_2^3}\right) = m\frac{d^2x_c}{dt^2} \quad (8)$$

$$-c_z z_c - \frac{1}{2}\left(\frac{q_2 z_1}{r_1^3} - \frac{q_1 z_2}{r_2^3}\right) = m\frac{d^2z_c}{dt^2} \quad (9)$$

Equations (6-9) comprise the set of equations necessary for the analytical method employed in this research. In the above, the two experimentally controllable parameters are the horizontal and vertical confinement, $c_x$ and $c_z$. The interaction potential between the two grains is characterized by setting the image charge magnitudes and displacements, $q_i$ and $l_i$, respectively.

Only the homogeneous forms of (6-9) are needed to investigate the equilibrium positions of the system. For any arbitrarily chosen $x$ and $z$, $c_x$ and $c_z$ can be determined through (6-7) and $x_c$ and $z_c$ can in turn be obtained from (8-9). In other words, the equilibrium positions of the pair as well as the required experimentally controllable parameters $c_x$ and $c_z$ are completely determined once $x$ and $z$ are specified. Thus a complete mapping of $c_x$, $c_z$, $x_c$ and $z_c$ on a two dimensional $x$-$z$ space of all possible equilibrium positions can be found using (6-9). Once equilibrium positions are identified, the non-homogeneous versions of (6-9) can be used to obtain the normal oscillation mode frequencies through standard linear stability analysis. Equilibrium positions having positive normal mode frequencies will be stable while negative mode frequencies represent instabilities in the system, driven by the corresponding mode. In this research, only the two cases most commonly observed in experiment, particle pairs aligned horizontally or vertically, are considered. The equilibrium positions, normal modes, and instabilities are all identified employing the above method and discussed separately below.

## III. HORIZONTAL PAIRS

The symmetry of horizontally aligned dust particle pairs allows analysis through the use of two image charges equal in magnitude and displacement, $q = q_1 = q_2$ and $l = l_1 = l_2$. To investigate the equilibrium positions of the system, the homogeneous form of (6-9) are used as described above. In this case, the vertical separation $z$ is equal to zero; therefore the equilibrium positions as well as the required confinement $c_x$ are determined by $x$ alone, and $c_x$ as a function of $x$ can be obtained from (6) (Fig. 3 a). On the other hand, in this case (7) can be satisfied for arbitrary $c_z$, making it an arbitrary input parameter.

The COM positions $x_c$ and $z_c$ can be calculated from (8, 9) for a specified equilibrium position, and the normal vibrational mode frequencies can be obtained from the non-homogeneous versions of (6-9) employing linear stability analysis.

The normal mode frequencies resulting from this approach, as a function of $x$ and normalized by $c_0 = Q/l_1^3$, are shown in Fig. 3. In this case, the value of $c_z$ was chosen to be 12.5 so that for a typical dust particle charge (i.e. thousands of elementary charges), the vertical resonance frequency of the system would be on the order of tens of Hertz, matching published experimental values [26]. A normalized image particle charge, $q = 2/3$ is assumed ($q$ can vary between 0 and 1, corresponding to a zero charge on the image particle to a charge on the image particle equal that on the dust particle).

Four modes exist for a specified value of $x$ ($c_x$). The horizontal and vertical frequencies corresponding to COM motion, $\omega_{H1}^2$ and $\omega_{V1}^2$, are equal to the horizontal and vertical confinements, $c_x$ and $c_z$ (Fig. 3a). As expected, the interparticle separation distance $x$ increases as the horizontal confinement $c_x$ becomes weaker.

The mode frequencies corresponding to the horizontal and vertical relative motion, $\omega_{H2}^2$ and $\omega_{V2}^2$, are also shown in Fig. 3a. Both are on the same order of magnitude as $c_z$, implying experimentally they should be observed to be a few $10^{th}$ of Hz. The frequency of the relative horizontal oscillation is large for small $x$ (large $c_x$) and approaches a limiting value of zero as $x$ increases ($c_x$ decreases). The frequency of the relative vertical oscillation determines the overall stability of the system. For very small $c_x$, $\omega_{V2}^2 \approx c_z$. However, as $c_x$ increases, $\omega_{V2}^2$ decreases and at some critical point, $c_{xcr}$, becomes negative, causing the horizontal pair to become unstable. This instability driven by the relative vertical oscillation is in agreement with previous theoretical work [14-17] and has been observed experimentally [21-23]. The critical value $c_{xcr}$ and the ratio $c_{xcr}/c_z$ are dependent on both $q$ and $c_z$ as shown in Fig. 4. (They are also dependent on $l$ and the dust particle charge $Q$ since $c_z$ is normalized by $c_0 \equiv Q/l^3$). When there is no wakefield ($q = 0$), the instability occurs for $c_{xcr}/c_z = 1$. With an increase in the wakefield ($q$ increases) the instability occurs for smaller values of horizontal confinement, indicating that the wakefield destabilizes the system. The weaker the vertical confinement $c_z$, the more dramatic this destabilization becomes. These results are again in agreement with both previous theoretical [14-17] and experimental [21-23] efforts.

For a typical experiment environment such as [21], the transition occurs at $c_{xcr}/c_z < 0.5$, where $c_z$ is around 0.01 when $l$ is a few hundred of microns, the same order as the interparticle distance. From Fig.4 this implies that the image charge (space charge) is very close in magnitude to the dust particle charge. From the threshold value alone $q$ and $l$ can not be determined at the same time. However, when combined with the spectrum data as shown in Fig.3, $q$ and $l$ should be measurable. This will be discussed in detail in an upcoming paper.

A comparison of Figs. 3 (a) and (b) show that the relative mode frequencies $\omega_{H2}^2$, $\omega_{V2}^2$ are the same as $\omega_H^2$, $\omega_V^2$, the horizontal and vertical mode frequencies when the COM motion is not considered, implying that in this case, the relative motion is decoupled from the COM motion, agreeing with previous research [15].

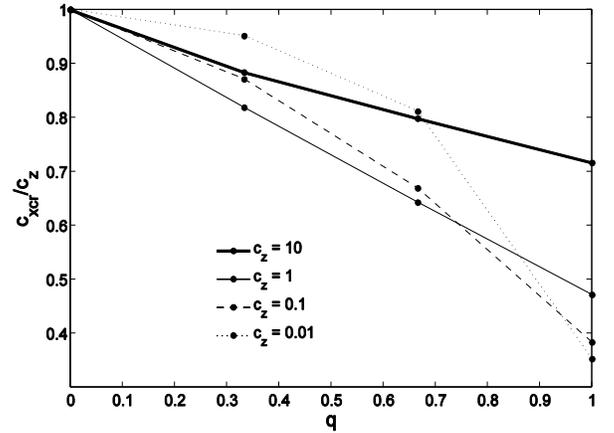

Fig. 4. Instability threshold $c_{xcr}/c_z$ for a horizontal dust particle pair as a function of $q$ for four different values of $c_z$.

## IV. VERTICAL PAIRS

A primary characteristic of vertically aligned dust particle pairs is that the two image charges are no longer equal. In fact, it can be shown from (7-10) that a vertically aligned system will never be stable for equal image charges. To take this into account requires the upper particle to have an image charge while the lower particle does not, i.e. $q = q_1$, $q_2 = 0$; $l = l_1$, $l_2 = 0$. This agrees with experiment [19, 20] where it has been found that the wake force generated by the top particle impacts the bottom particle, but not the other way around. A limitation of this model is that when the interparticle distance is small, i.e. $z < l$, the ion cloud is modified in such a manner as to make the image particle model invalid. As a result, in this work we consider only cases where $z > l$.

For a vertically aligned pair, $x = 0$; therefore, $c_z$ is a function of $z$ only and the horizontal confinement $c_x$ becomes an arbitrary input parameter. Fig. 5 shows the normal mode frequencies for $q = 2/3$ and $c_x = 10$ (chosen to be of the same order as $c_z$ used previously for analysis of a horizontal pair). As shown, the frequencies for the horizontal modes $\omega_{H1}^2$, $\omega_{H2}^2$ are approximately 10 (Fig. 5 a) while the vertical mode frequencies $\omega_{V1}^2$, $\omega_{V2}^2$ are close to 0 (Fig. 5 b). Again, $\omega_{H1}^2 = c_x$, $\omega_{V1}^2 = c_z$, and the interparticle distance $z$ increases as the confinement $c_z$ decreases, as expected. In this case, the vertical relative mode $\omega_{V2}^2$ determines the overall stability of the system. For small $c_z$, (or large $z$), $\omega_{V2}^2$ is low. As $c_z$ increases, $\omega_{V2}^2$ also increases, reaching a maximum value before dropping sharply. When $c_z$ becomes greater than the threshold value $c_{zcr}$, $\omega_{V2}^2 < 0$, and the pair becomes unstable driven by the vertical relative mode.

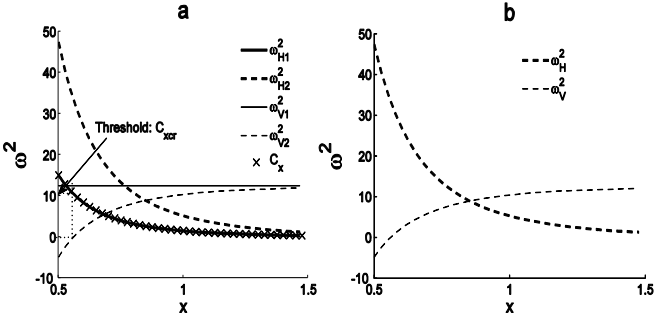

Fig. 3. The normal mode frequencies as a function of $x$ for a horizontal dust particle pair having $q = 2/3$ and $c_z = 12.5$, (a) with and (b) without the COM motion taken into account. $\omega_{H1}$, $\omega_{H2}$, $\omega_{V1}$, $\omega_{V2}$, $\omega_H$, $\omega_V$, are as defined in the text.

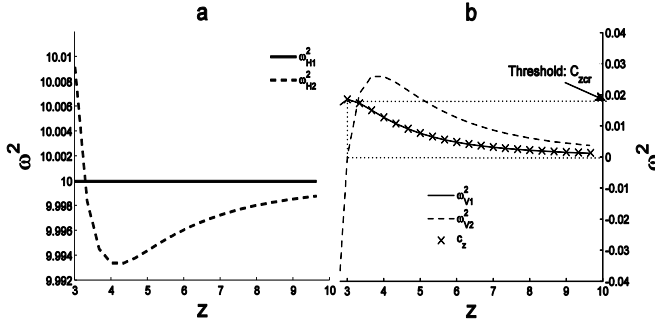

Fig. 5. Normal mode frequencies (a) $\omega_{H1}^2$, $\omega_{H2}^2$ and (b) $\omega_{V1}^2$, $\omega_{V2}^2$ as a function of $z$ for a vertically aligned pair with $q = 2/3$ and $c_x = 10$. Note the difference in vertical scale.

It is interesting to note that vertical pairs differ from horizontal pairs in that they exhibit a second type of instability which is driven by the *horizontal* relative mode. As $c_x$ decreases beyond a critical value (which is a function of $q$ only), $\omega_{H2}^2$ becomes negative within a specified range of $c_z$ ($c_{zcr1} < c_z < c_{zcr2}$) and the system becomes unstable (Fig. 6 b). As shown, the shape of the frequency curves are also independent of $c_x$, although the magnitude of $\omega_{H1}^2$, $\omega_{H2}^2$ vary as $c_x$ varies. Thus, for a specified $q$, the system remains stable within the same range $c_z > c_{zcr}$ as long as $c_x$ is larger than the critical value (Fig. 6).

It can also be seen in Fig.6 that, in contrast to the relative modes for horizontally aligned pairs, $\omega_{H2}^2$, $\omega_{V2}^2$ are much smaller in magnitude (around a few Hertz) for typical experimental environments. This has been observed in previous experiments [20] and also in our lab in CASPER, which will be discussed in a paper currently under preparation.

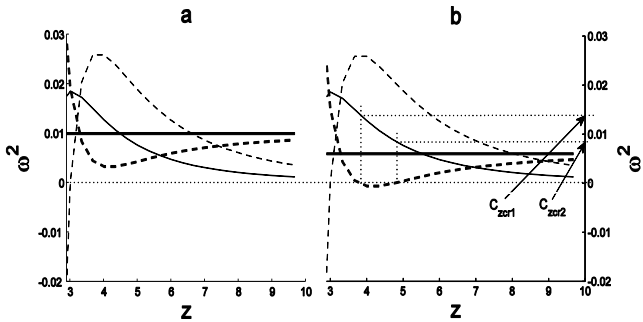

Fig. 6. Normal mode frequencies as a function of $z$ for a vertically aligned pair for (a) $c_x = 0.01$ (b) $c_x = 0.006$.

Finally, it can be seen that the COM does not coincide with the minimum of the potential well on the vertical direction for either horizontally or vertically aligned dust particle pairs, in agreement with previous theory [14, 16]. Fig. 7 quantitatively shows the COM height as a function of the interparticle distance. For a horizontal pair, the COM remains below the center of the potential well while for a vertical pair, the COM sits well above it. For both cases, the distance between the COM and the potential well center $|z|_c$ decreases as the interparticle distance increases (Fig. 7).

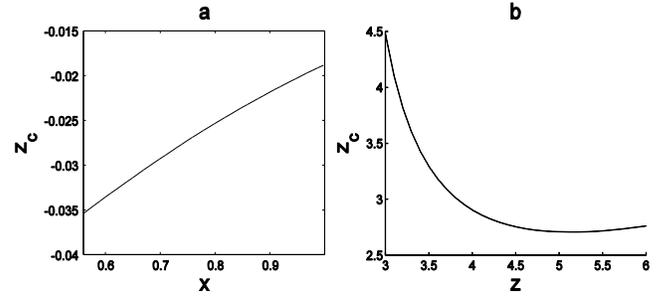

Fig. 7. COM heights for (a) horizontal ($c_z = 12.5$) and (b) vertical ($c_x = 10$) dust particle pairs as a function of interparticle distance $x(z)$. ($q = 2/3$).

## V. SUMMARY

The equilibrium positions, normal modes, and regions of instability for dust particle pairs in a complex plasma have been investigated in detail employing an analytical method, taking into account the electric-field induced plasma modification (wakefield) around the dust particles. The wakefield was modeled using a simple but effective image charge method. Newton's equations were employed assuming a non-Hamiltonian interaction as well as a parabolic external confinement in both the horizontal and vertical directions and the center of mass motion for the system was included.

Two special cases, dust particle pairs aligned horizontally and vertically, were investigated. Horizontal pairs were assumed to have identical image charges, while only the upper particles in vertical pairs were allowed to have an image charge, providing a mechanism for examination of the anisotropic force in the vertical direction. For any specified set of controlling parameters $c_x$ (horizontal confinement) and $c_z$ (vertical confinement) for which the pair remained in equilibrium, it was shown that four normal modes exist. Two of these correspond to the horizontal and vertical COM motion and have frequencies equal to $c_x$ and $c_z$. An additional two modes, decoupled from the COM modes, correspond to the horizontal and vertical relative motion and act as primary drivers for the instabilities.

Both horizontally and vertically aligned pairs can become unstable as $c_x$ or $c_z$ is varied. For a horizontally aligned pair, such system instability is induced through varying $c_x$, and driven by the vertical relative oscillation mode. For a vertical pair, two different types of instabilities exist; both are induced by varying $c_z$, but one is driven by the vertical relative oscillation mode, while the other is driven by the horizontal relative oscillation mode. As expected, the COM was shown to not coincide with the minimum of the potential well due to the interaction of the image charges, which are related to the non-Hamiltonian characteristic of the interaction. Future studies will examine these behaviors experimentally and compare them with current results, which should lead to a more quantitative determination of the pair interaction including the wakefield.

The method developed above may also be used to obtain the mode spectra for oblique equilibriums, such as those which can be creted by the dependence of the image charge on the height above the lower electrode (i.e., for a non-harmonic confinement). This will be addressed in a paper currently under preparation.

## REFERENCES

[1] U. Konopka, G. E. Morfill and L. Ratke, *Phys. Rev. Lett.*, vol. 84, pp. 891-894, 1999.
[2] R. Kompaneets, U. Konopka, A. V. Ivlev, V. Tsytovich and G. Morfill, *Physics of Plasmas*, 14, 052108, 2007.
[3] S. V. Vladimirov and M. Nambu, Phys. Rev. E 52, 2172 (1995).
[4] F. Melandsø and J. Goree, Phys. Rev. E 52, 5312 (1995).
[5] J. H. Chu and Lin I, Phys. Rev. Lett. 72, 4009 (1994).
[6] Y. Hayashi and K. Tachibana, Jpn. J. Appl. Phys. 33, L804 (1994).
[7] H. Thomas, G. E. Morfill, V. Demmel, J. Goree, B. Feuerbacher, and D. Möhlmann, Phys. Rev. Letters 73, 652 (1994).
[8] A. Homann, et al.., Physics Letters A, 242, 173-180 (1998).
[9] S. Nunomura, et al.., Phys. Rev. Lett. 89, 35001(1-4) (2002).
[10] K. Qiao and T. W. Hyde, Physical Review E 68, 046403 (2003).
[11] K. Qiao and T. W. Hyde, Physical Review E 71, 026406 (2005).
[12] A. Barkan, R. L. Merlino, and N. D'Angelo, Phys. Plasmas 2, 3563 (1995).
[13] A. V. Ivlev, et al. Phys. Rev. Lett. 100, 095003 (2008).
[14] Martin Lampe, Glenn Joyce and Gurudas Ganguli, IEEE transactions on Plasma Science, 33, 57, 2005.
[15] S. V. Vladimirov and A. A. Samarian, Phys. Rev. E, 65, 046416, 2002.
[16] J. D. E. Stokes, A. A. Samarian and S. V. Vladimirov, Phys. Rev. E 78, 036402, 2008.
[17] A. A. Samarian and S. V. Vladimirov, Contrib. Plasma Phys. 49, 260-280, 2009.
[18] V. V. Yaroshenko, A. V. Ivlev, and G. E. Morfil, Phys. Rev. E 71, 046405 (2005).
[19] A. Melzer, Physica Scripta. T89, 33-36, 2001.
[20] G. A. Hebner and M. E. Riley, Phys. Rev. E 68, 046401, 2003.
[21] V. Steinberg, et al. Phys. Rev. Lett. 86, 4540 (2001).
[22] A. Melzer, V. A. Schweigert, and A. Piel, Phys. Rev. Lett. 83, 3194 (1999).
[23] A. A. Samarian, S. V. Vladimirov and B. W. James, Physics of Plasmas 12, 022103, 2005.
[24] Giovanni Lapenta, Physica Scripta. 64, 599-604, 2001.
[25] P. Ludwig, S. Kosse, and M. Bonitz, *Phys. Rev. E*, 71, 046403 (2005).
[26] E. B. Tomme, et al., Phys. Rev. Lett. 85, 2518 (2000).

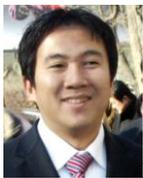
**Ke Qiao** received the B.S. degree in physics from Shandong University, Qingdao, China, and the Ph.D. degree in theoretical physics from Baylor University, Waco, TX. He is currently with Baylor University, where he is an Assistant Research Scientist at the Center for Astrophysics, Space Physics, and Engineering Research (CASPER). His research interests include structure analysis, waves and instabilities, and phase transitions in complex (dusty) plasmas.

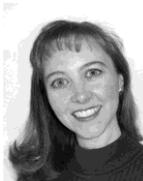
**Lorin Swint Matthews** was born in Paris, TX in 1972. She received the B.S. and the Ph.D. degrees in physics from Baylor University in Waco, TX, in 1994 and 1998, respectively. She is currently an Assistant Professor in the Physics Department and Senior Research Scientist at CASPER (Center for Astrophysics, Space Physics, & Engineering Research) at Baylor University. Previously, she worked at Raytheon Aircraft Integration Systems where she was the Lead Vibroacoustics Engineer on NASA's SOFIA (Stratospheric Observatory for Infrared Astronomy) project.

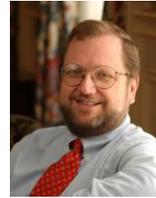
**Truell W. Hyde** received the B.S. in physics and mathematics from Southern Nazarene University and the Ph.D. in theoretical physics from Baylor University. He is currently at Baylor University where he is the Director of the Center for Astrophysics, Space Physics & Engineering Research (CASPER), a professor of physics and the Vice Provost for Research for the University. His research interests include space physics, shock physics and waves and nonlinear phenomena in complex (dusty) plasmas.